\documentclass[a4paper]{jpconf}
\usepackage{amsmath,amsfonts,amssymb}
\usepackage{graphicx}
\usepackage[colorlinks=true, allcolors=blue]{hyperref}
\begin{document} 
\title{Distributed Coordination for Multi-Vehicle Systems in the Presence of Misbehaving Vehicles}

\author{Dongkun Han}
\address{Department of Mechanical and Automation Engineering, The Chinese University of Hong Kong, Hong Kong SAR, China}
\ead{dkhan@mae.cuhk.edu.hk}

\author{Yijun Huang}
\address{Department of Mechanical and Automation Engineering, The Chinese University of Hong Kong, Hong Kong SAR, China}
\ead{yjhuang@link.cuhk.edu.hk}

\author{Hejun Huang}
\address{Department of Aerospace Engineering, University of Michigan, Ann Arbor, U.S.}
\ead{hejun@umich.edu}

\author{Tianrui Fang}
\address{Department of Mechanical and Automation Engineering, The Chinese University of Hong Kong, Hong Kong SAR, China}
\ead{1155210076@link.cuhk.edu.hk}

\begin{abstract}
The coordination problem of multi-vehicle systems is of great interests in the area of autonomous driving and multi-vehicle control. This work mainly focuses on multi-task coordination problem of a group of vehicles with a bicycle model and some specific control objectives, including collision avoidance, connectivity maintenance and convergence to desired destinations. The basic idea is to develop a proper Lyapunov-like barrier function for all tasks and a distributed controller could be built in the presence of misbehaving vehicles. Control protocols are provided for both leader vehicle and follower vehicles. The simulation results demonstrate the effectiveness of proposed method. 
\end{abstract}

\section{Introduction}
\label{sec:intro}  % \label{} allows reference to this section

As autonomous driving technology becomes increasingly prevalent, the focus on multi-agent systems has intensively grown in recent years \cite{li22tal}. One of the main challenges in autonomous driving is coordinating different control objectives of multiple vehicles, both on a global and individual level. Recent surveys have highlighted the introduction of various distributed coordination and control problems \cite{dorri18access,ref3}. 

Stability and safety are both significant properties in multi-agent systems. To guarantee these two properties, by mixing the Lyapunov stability theory and control barrier function, the concept of Lyapunov-like barrier functions has been introduced for multi-task coordination problems \cite{ref1,ames16tac}. This method has been successfully implemented in mechanical system design \cite{rauscher16iros}, robotic grasping \cite{cortez19tcst}, adaptive cruise control \cite{dey15review}, and lane-keeping maneuvering \cite{xu17tase}.  

Unicycle model is commonly used to consider a group of agents with respective desired objectives under constraints like collision avoidance and connectivity maintenance, which can drive multi-agent to desired point under various constraints \cite{ref7}. Unicycle kinematic model can be described by:
\begin{equation}
\label{sys1}
    \left\{
    \begin{array}{rcl}
    \dot{x_i}=&u_i\cos\theta_i,
    \vspace{0.5em}\\
    \dot{y_i}=&u_i\sin\theta_i,
    \vspace{0.5em}\\
    \dot{\theta_i}=&\omega_i, 
    \end{array}
\right.
\end{equation}
where $[x_i,y_i,\theta_i]$ is the configuration vector of agent $i$, $u_i$ is the linear velocity and $\omega_i$ is the angular velocity as control inputs. Unicycle model has been widely adopted in robotics and unmanned systems due to its simplicity and applicability. However, regards to autonomous driving, it is not easy to depict the relationship and dynamics between front wheel and back wheel of vehicles. Thus, it cannot provide proper and realistic advisory parameters in controller design.  

In contrast to unicycle models, a bicycle model is investigated in this work as follows.

%\section*{\romannumeral2. System Modeling}
%\section*{SYSTEM MODELING}
\subsection{Model of Autonomous Vehicles}

\begin{figure}[thpb]
\label{fig1}
\centering
\includegraphics[width=0.4\linewidth]{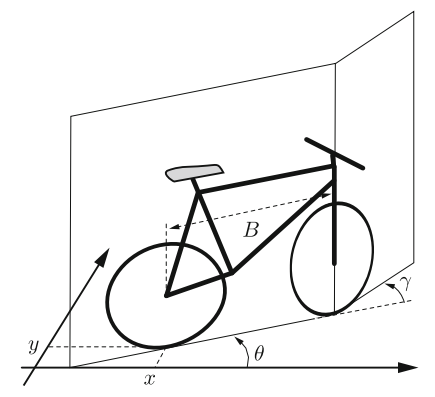}
\caption{Bicycle model schematic}
\end{figure}

A typical bicycle model is shown in Figure \ref{fig1}. It is worthy noting that the angular velocity $\omega$ of the front wheel and the linear velocity $u$ of the rear wheel are both provided by control inputs. The angle of the rear wheel denotes as $\theta$ which is influenced both by the inputs $u$ and $\omega$. If we consider each vehicle as an agent, the kinematic bicycle model for each agent $i\in \{1,2,...,N\}$ is given as:

\begin{equation}
\label{sys2}
\left\{
    \begin{array}{rcl}
    \dot{x_i}=&u_i\cos\theta_i,
    \vspace{0.5em}\\
    \dot{y_i}=&u\sin\theta_i,
    \vspace{0.5em}\\
    \dot{\theta_i}=&\frac{u_i}{B}\tan \gamma_i,
    \vspace{0.5em}\\
    \dot{\gamma_i}=&\omega_i,
    \end{array}
\right.
\end{equation}
where we consider the location of the rear wheal $(x,y)$ as the location of the agent, $B$ is the distance between the front wheel and the rear wheel, $\theta$ is the angle of the bicycle frame with respect to the x-axis, and $\gamma$ is the angle of the front wheel with respect to the bicycle frame.

\begin{figure}[thpb]
\label{fig2}
\centering
\includegraphics[width=0.5\linewidth]{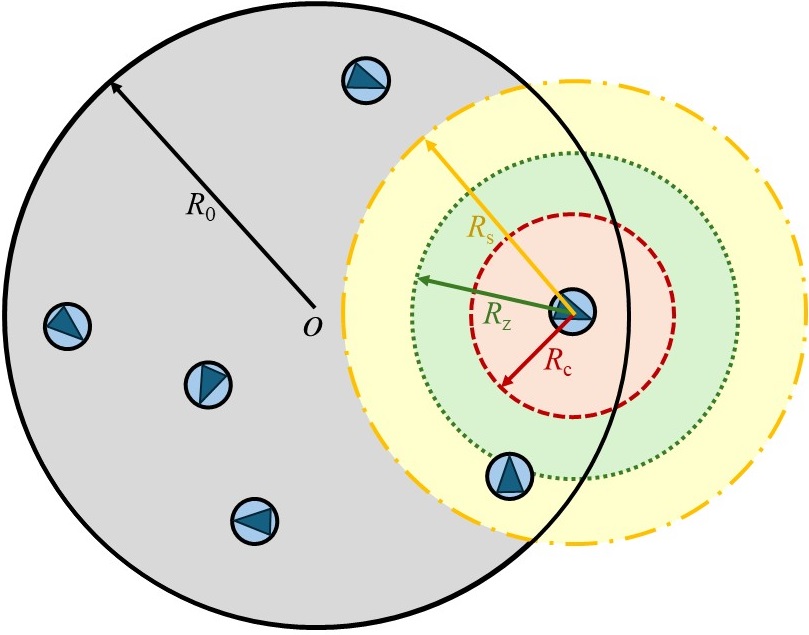}
\caption{Configuration illustration for the leader agent and follower agents.}
\end{figure}

Each agent can be denoted by three distinct circles, whose radius are $R_s$,$R_z$,$R_c$ from largest to smallest as shown in Fig. 2. $R_s$ is the radius of sensing region of every agent, i.e., each follower agent can measure the position of other agents within the distance $d_{ij} \leq R_s$. $R_c$ is the radius of safety region, i.e., the movement speed and direction of other agents within safety region would be sent to the corresponding agent and prevent any agent in this region. The radius of circular region $R_z$ denotes the region in which the collision avoidance objective is active for agent $i$ and it is used in the design of followers' controller in section \uppercase\expandafter{\romannumeral4}.

\subsection{Problem Statement}
In this study, we focus on a network comprising $N$ (where $N > 1$) agents. The objective is to develop a distributed controller that can guide these agents towards their respective target positions in a two-dimensional coordinate space. It is crucial to ensure that the agents avoid collisions with both other agents and detectable obstacles. Furthermore, the communication network's connectivity must be maintained, meaning that all agents are confined to a circular region denoted as $O$ with a radius of $R_0$. Additionally, we aim to ensure that all three objectives can be achieved even in the presence of misbehaving agents.

The objective of this study is to develop a distributed controller for multi-agent systems, including the leader and the followers, that ensures their convergence to desired destinations under specific constraints. The distributed controller design must address three main requirements: Collision avoidance, connectivity maintenance, and convergence to desired destinations. To satisfy these constraints, a new type of barrier functions is developed. The system comprises a leader agent and follower agents, each driven by their respective controllers to converge towards their desired destinations. In addition, the controller should be capable of achieving all these goals even in extreme cases, such as when there are misbehaving agents.

\section{Design of Lyapunov-like Barrier Functions}
Barrier functions are continuous functions that tend towards infinity as they approach the boundaries of feasible regions defined by constraints such as connectivity maintenance and collision avoidance. The concept of recentered barrier functions has been previously introduced to ensure convergence of the barrier function towards the destination if the entire system is capable of converging \cite{ref1}.

%\subsection*{A. Connectivity maintenance}
\subsection{Connectivity maintenance}
All agents cannot converge to their desired destination if there are some agents isolated and cannot communicate with each other. An intuitive way to preserve connectivity is to keep all $N$ agents in a circular region $O$ located at $\mathbf{r_0}=[x_0\ y_0]^T$ with a radius of $R_0$, such that all the followers could be communicated by the leader. For any agent $i$, the distance between $\mathbf{r_i}=[x_i\ y_i]^T$ and $\mathbf{r_0}=[x_0\ y_0]^T$ is denoted as $d_{i0}$, which should remain less or equal to R ($R=R_0-r_a$, where $r_a$ is the radius of each agent). This condition can be expressed as follows:
\begin{equation}
    c_{i0}(\mathbf{r_i})=R-||\mathbf{r_i}-\mathbf{r_0}||=R-d_{i0}\ge 0.
\end{equation}
For every single agent $i$, the above constraint should always hold to maintain the connectivity of network. Then, we can define the barrier function as:
\begin{equation}
    b_{i0}(\mathbf{r_i})=-\ln(c_{i0}(\mathbf{r_i})),
\end{equation}
where $b_{i0}(\mathbf{r_i})\to+\infty$ when $c_{i0}(\mathbf{r_i})\to0$, from the definition of $b_{i0}(\mathbf{r_i})$, we can then define the recentered barrier function for \begin{equation}
    r_{i0}(\mathbf{r_i})=b_{i0}(\mathbf{r_i})-b_{i0}(\mathbf{r_{id}})-(\nabla{b_{i0}|_{\mathbf{r_{id}}}})^T(\mathbf{r_i}-\mathbf{r_{id}}),
\end{equation}
where $\mathbf{r_{id}}=[x_{id}\ y_{id}]^T$ is the destination of agent $i$ and $\nabla{b_{i0}}$ is the gradient vector of the function $b_{i0}(\mathbf{r_i})$, i.e. we have $\nabla{b_{i0}}=[\frac{\partial b_{i0}}{\partial x_i}\ \frac{\partial b_{i0}}{\partial y_i}]$, and $(\nabla{b_{i0}|_{\mathbf{r_{id}}}})^T$ is the transpose of the gradient vector at the destination $\mathbf{r_{id}}=[x_{id}\ y_{id}]^T$ of agent $i$.

Some properties of the recentered barrier function $r_{i0}(\mathbf{r_i})$ can be discussed here. First, $r_{i0}(\mathbf{r_i})$ tends to 0 if and only if the position of agent $i$ is at its desired position, i.e. $\mathbf{r_i}=\mathbf{r_{id}}$. $r_{i0}(\mathbf{r_i})$ tends to positive infinity if and only if $c_{i0}$ tends to 0, i.e. the position of agent $i$ approaches to the boundary of connectivity region $O$. For each agent $i\in \{1,2,...,N\}$ we can define the Lyapunov-like barrier function as:
\begin{equation}
    V_{i0}(\mathbf{r_i})=(r_{i0}(\mathbf{r_i}))^2
\end{equation}
From the above definition, it is obvious that $V_{i0}(\mathbf{r_i})$ is a non-negative function.

\subsection{Collision Avoidance}
For the leader agent ($i=1$), it only needs to consider physical obstacles detected in its sensing region, while for the follower agents $i\in \{2,3,...,N\}$, they should avoid collision with both physical obstacles and other agents $j\in \{1,2,...,N\},j\ne i$. Thus, for agent $i\in \{2,3,...,N\}$, all the other agents are considered as physical obstacles. Regarding collision avoidance, we should ensure that the distance of agent $i$ and agent $j$ not less than the minimum separation distance $d_s$, which can be written as:
\begin{equation}
    c_{ij}(\mathbf{r_i},\mathbf{r_j})=||\mathbf{r_i}-\mathbf{r_j}||^2-ds^2=(x_i-x_j)^2+(y_i-y_j)^2-d_s^2 \ge 0
\end{equation}
For each agent $i\in \{2,3,...,N\}$, we have $(N-1)\times (N-2)$ constraints of collision avoidance. Similar to the part in the section of connectivity maintenance, we can define the barrier function as:
\begin{equation}
    b_{ij}(\mathbf{r_i},\mathbf{r_j})=-\ln(c_{ij}(\mathbf{r_i},\mathbf{r_j}))
\end{equation}
where $b_{ij}(\mathbf{r_i},\mathbf{r_j})\to +\infty$ when $c_{ij}(\mathbf{r_i},\mathbf{r_j})\to 0$. The recentered barrier function $q_{ij}$ then can be defined according to $b_ij$:
\begin{equation}
    q_{ij}(\mathbf{r_i},\mathbf{r_j})=b_{ij}(\mathbf{r_i},\mathbf{r_j})-b_{ij}(\mathbf{r_{id}},\mathbf{r_j})
\end{equation}
where the function $q_{ij}(\mathbf{r_i},\mathbf{r_j})$ tends to 0 if and only if the position of agent $i$ is at its desired position, i.e. $\mathbf{r_i}=\mathbf{r_{id}}$ and $q_{ij}(\mathbf{r_i},\mathbf{r_j})$ tends to positive infinity if and only if the distance between agent $i$ and agent $j$ tends to $d_s$. Then we can define $V_{ij}'$ as:
\begin{equation}
    V_{ij}'(\mathbf{r_i},\mathbf{r_j})=(q_{ij}(\mathbf{r_i},\mathbf{r_j}))^2
\end{equation}
which is positive definite. Considering the sensing region $R_s$ and the collision avoidance region $R_z$, we can build the Lyapunov-like barrier function $V_{ij}$ from  $V_{ij}'$ as:
\begin{equation}
    V_{ij}(\mathbf{r_i},\mathbf{r_j})=\sigma_{ij}V_{ij}'(\mathbf{r_i},\mathbf{r_j})
\end{equation}
where $\sigma_{ij}$ can be defined as:
$$
    \sigma_{ij}=\begin{cases}
        1,&\text{if}\ d_s\le d_{ij}\le R_z
        \vspace{0.5em}\\
        Ad_{ij}^3+Bd_{ij}^2+Cd_{ij}+D,&\text{if}\ R_z<d_{ij}<R_s
        \vspace{0.5em}\\
        0,&\text{if}\ d_{ij}\ge R_s
    \end{cases}
$$
where $\ A=-\frac{2}{(R_z-R_s)^3}$,$\ B=\frac{3(R_z+R_s)}{(R_z-R_s)^3}$,$\ C=-\frac{6R_zR_s}{(R_z-R_s)^3}$,$\ D=-\frac{R_s^2(3R_z-R_s)}{(R_z-R_s)^3}$. In this way, we can ensure collision avoidance for agent $i$ when agent $j$ is within sensing region and out of collision avoidance region.

\subsection{Combination of Constraints}
From the above two subsections, we have obtained two different Lyapunov-like barrier functions $V_{i0}$ and $V_{ij}$ built for connectivity maintenance and collision avoidance, respectively. Then we need one single Lyapunov-like barrier function for each agent which can satisfy connectivity maintenance, collision avoidance and convergence objectives. Here we propose a method to combine $V_{i0}$ and $V_{ij}$ together as a single Lyapunov-like barrier function $v_i$ like:
\begin{equation}
    v_i=\left((V_{i0})^\delta+\sum_{j=1,j\ne i}^{N}(V_{ij})^\delta\right) ^{\frac{1}{\delta}}=\left(\sum_{n\in M}(V_{ij})^\delta\right)^{\frac{1}{\delta}}
\end{equation}
where $M=\{0,1,...N\}$, $\delta\in [1,+\infty]$. The Lyapunov-like barrier function tends to 0 if and only if all $V_{i0}$ and $V_{ij}$ tends to 0. From section A and B, we know that only when the position of agent $i$ is at its desired position, i.e. $\mathbf{r_i}=\mathbf{r_{id}}$, $V_{i0}$ and $V_{ij}$ could become 0. Meanwhile, $v_i$ tends to positive infinity when at least one of $V_{in}$ tends to positive infinity. When the agent $i$ approaches to the boundary of connectivity region $O$ or tends to have collision with other agents, $v_i$ will tend to $+\infty$. To normalize the Lyapunov-like barrier function, we define $V_i$ as:
$$
    V_i=\frac{v_i}{1+v_i}=\frac{\left(\sum_{n\in M}(V_{ij})^\delta\right)^{\frac{1}{\delta}}}{1+\left(\sum_{n\in M}(V_{ij})^\delta\right)^{\frac{1}{\delta}}}
$$
whose value is 1 when $v_i$ tends to positive infinity. The value turns to be 0 when $v_i$ tends to 0. $V_i$ is positive definite and equals to 0 if and only if agent $i$ is at its destination.

\section{Main Results: Controller Design}

% 3 obstacles, agent 10-12-15

In this Section, we will design the distributed controller for multi-vehicle systems with and without the presence of misbehaving vehicle. All agents stay within the connectivity region $O$ all the time with reliable wireless information exchange. The leader agent $j=1$ is expected to be controller to its desired position while for all the follower agents $j\ne 1$ need to converge to its desired position avoiding colliding with other agents and maintaining connectivity. We first consider the case that there is no misbehaving agents and the distributed controller is always reliable.

\subsection{Controller for the leader agent}

Based on the Lyapunov-like barrier function constructed in the previous section, we can now propose our main results regarding the distributed controller design:

\noindent\textbf{Theorem 1.} Considering the multi-vehicle systems depicted by a bicycle model \ref{sys2}, the leader agent can converge to its desired position with connectivity maintenance and collision avoidance if it is driven by the distributed controller:
\begin{equation}
    \begin{array}{cc}
        u_1=&k_1\tanh{(||\mathbf{r_1}-\mathbf{r_{1d}}||)}
        \vspace{0.5em}\\
        \omega_1=& \displaystyle \frac{(B\lambda_1 \dot{\phi}_1-\lambda_1u_1\tan{\gamma_1}+B\ddot{\phi}_1)u_1+(\dot{\phi}_1-\lambda_1\theta_1+\lambda_1\phi_1)\dot{u}_1}{u_1^2+B(\dot{\phi}_1-\lambda_1 \theta_1+\lambda_1\phi_1)^2}\\
    \end{array}
\end{equation}
where $\tanh$ is the hyperbolic tangent function, $k_1>0, \lambda_1>0$, and $\phi_1$ is the orientation of the negated gradient of the Lyapunov-like barrier function, i.e. $\phi_1={\rm atan2}\left(-\frac{\partial V_1}{\partial x_1},-\frac{\partial V_1}{\partial y_1}\right)$.

\emph{Proof}: In this proof, we can show that a bicycle system can be transformed into two special unicycle systems. First, we will propose a distributed controller for an unicycle system (in Part 1). Then, a distributed controller will be built for the concerned bicycle system (in Part 2). 

 Part 1: A proof will be given that multi-vehicle system can converge to the desired posit ion with connectivity maintenance and collision avoidance for the unicycle system\cite{ref1} under the following control protocol:
\begin{equation}
    \begin{array}{cc}
        u_1=&k_1\tanh{(||\mathbf{r_1}-\mathbf{r_{1d}}||)}
        \vspace{0.5em}\\
         \omega_1=&-\lambda_1(\theta_1-\phi_1)+\dot\phi_1.\\
    \end{array}
\end{equation}

For unicycle system, $\omega_1=\dot\theta_1$, we have:
\begin{equation}
    \frac{1}{\lambda_1}\dot\theta_1=-(\theta_1-\phi_1)+\frac{1}{\lambda_1}\dot\phi_1
    \Rightarrow
    \theta_1-\phi_1=-\frac{1}{\lambda_1}(\dot\theta_1-\dot\phi_1).
\end{equation}
Then, we define $\varepsilon_1=\frac{1}{\lambda_1},\eta_1=\theta_1-\phi_1$, where $\varepsilon_1>0$ and,$\ \dot\theta_1-\dot\phi_1=\frac{d\eta_1}{dt}$. It results in:
\begin{equation}
    -\eta_1=\varepsilon_1\frac{d\eta_1}{dt}
    \Rightarrow
    \eta_1=e^{-\varepsilon_1t}
\end{equation}

Thus, $\varepsilon_1$ exponentially converges to $0$, i.e. $\theta_1$ exponentially converges to the orientation of the negated gradient of Lyapunov-like barrier function $\phi_1$.

For leader agent $j=1$, we have obtained a positive definite Lyapunov-like barrier function $V_1$ derived from Section \uppercase\expandafter{\romannumeral3}. In the following, we will check whether $\dot V_j$ is negative such that the convergence can be achieved. Let us define the gradient vector of Lyapunov-like barrier function as $\zeta_1=\left[\frac{\partial V_1}{\partial x_1},\frac{\partial V_1}{\partial y_1}\right]^T$. Thus, we can derive that:
\begin{equation}
    \dot V_1=\left[\frac{\partial V_1}{\partial x_1},\frac{\partial V_1}{\partial y_1}\right]\left[\begin{array}{cc}
         \dot x_1\\
         \dot y_1
    \end{array}\right]=\left(\frac{\partial V_1}{\partial x_1}\cos{\theta_1}+\frac{\partial V_1}{\partial y_1}\sin{\theta_1}\right)u_1.
\end{equation}
Put the definition of $u_1$ into the formula, we have:
\begin{equation}
    \dot V_1=k_1\left(\frac{\partial V_1}{\partial x_1}\cos{\theta_1}+\frac{\partial V_1}{\partial y_1}\sin{\theta_1}\right)\tanh{(||\mathbf{r_1}-\mathbf{r_{1d}}||)}
\end{equation}
Then we define the gradient vector $\mathbf{\zeta_1}=\left[\frac{\partial V_1}{\partial x_1}\ \frac{\partial V_1}{\partial y_1} \right]^T$,and $\phi_1=atan2\left(-\frac{\partial V_1}{\partial x_1},-\frac{\partial V_1}{\partial y_1}\right)$, we have:
\begin{equation}
    \frac{\partial V_1}{\partial x_1}=-||\mathbf{\zeta_1}||\cos{\phi_1},\ 
    \frac{\partial V_1}{\partial y_1}=-||\mathbf{\zeta_1}||\sin{\phi_1}.
\end{equation}
As we have shown that $\theta_1$ exponentially converges to the orientation of the negated gradient of Lyapunov-like barrier function $\phi_1$. We can simply assume that the orientation $\theta_1$ is controlled much faster than the translation, so that we can simplify $\dot V_1$ by putting $\theta_1=\phi_1$ into the formulation:
\begin{equation}
    \dot V_1=-k_1(||\mathbf{\zeta_1}||\cos^2{\phi_1}+||\zeta_1||\sin^2{\phi_1})\tanh{(||\mathbf{r_1}-\mathbf{r_{1d}}||)}
    =-k_1||\mathbf{\zeta_1}||\tanh{(||\mathbf{r_1}-\mathbf{r_{1d}}||)}
\end{equation}
where $k_1>0$, $\ ||\mathbf{\zeta_1}\ge 0||$ and $\tanh{(||\mathbf{r_1}-\mathbf{r_{1d}}||)}\ge 0$. Hence, $\dot V_1$ is not greater than 0, i.e. $\dot V_1\le 0$ and equals to 0 if and only if agent $j=1$ is at its desired position, i.e. $\mathbf{r_1}=\mathbf{r_{1d}}$.

Part 2: We can divide one single bicycle system into two unicycle systems. Assuming $\widetilde\omega_1$ is the controller for unicycle system. And $\omega_1$ is for bicycle system, we can derive the orientation of agent 1 at $t_1=t_0+dt$ from $t_0$:
\begin{equation}
    \begin{array}{cc}
        \widetilde\theta_1(t_0+dt)=&\widetilde\theta_1(t_0)+\widetilde\omega_1dt
        \vspace{0.5em}\\
        \gamma_1(t_0+dt)=&\gamma_1(t_0)+\omega_1dt
        \vspace{0.5em}\\
        \theta_1(t_0+dt)=&\theta_1(t_0)+\frac{u_1}{B}\tan{\gamma_1(t_1)}dt
    \end{array}
\end{equation}
where $\widetilde\theta_1$ is the orientation of unicycle, and $\theta_1$ is the orientation of of bicycle and $\gamma_1$ is the orientation of the front wheel. Then we can derive the controller of bicycle $\omega_1$ from the controller of unicycle $\widetilde\omega_1$:
\begin{equation}
    \widetilde\omega_1=\frac{u_1}{B}\tan{(\gamma_1(t_0)+\omega_1dt)}
    \Rightarrow
    \omega_1=\frac{\arctan{\frac{B\widetilde\omega_1}{u_1}}-\gamma_1}{dt}
\end{equation}

Replace $\widetilde\omega_1$ that we derived in Part 1 for unicycle systems, it yields that:
\begin{equation}
\begin{array}{rcl}
    \omega_1 =\dot\gamma &=& \displaystyle \frac{\arctan\left(\frac{B(-\lambda_1(\theta_1-\phi_1)+\dot\phi_1)}{u_1}\right)-\gamma_1}{dt}   \\
    &=& \displaystyle \frac{(B\lambda_1 \dot{\phi}_1-\lambda_1u_1\tan{\gamma_1}+B\ddot{\phi}_1)u_1+(\dot{\phi}_1-\lambda_1\theta_1+\lambda_1\phi_1)\dot{u}_1}{u_1^2+B(\dot{\phi}_1-\lambda_1 \theta_1+\lambda_1\phi_1)^2}
\end{array}
\end{equation}
in which $u_1=k_1\tanh{(||\mathbf{r_1}-\mathbf{r_{1d}}||)}$. Through the same way for unicycle systems, one could get $\dot V_1\le 0$ and equals to 0 if and only if agent $j=1$ is at its desired position, i.e. $\mathbf{r_1}=\mathbf{r_{1d}}$. Thus, it completes the proof. \hfill $\square$

\subsection{Controller for the follower agents}
In this subsection, we will show how we design the distributed controller for the follower agents which can converge to their goal destination $\mathbf{r}_{jd}$ with connectivity maintenance and collision avoidance. 

\noindent\textbf{Theorem 2.} For any follower agent $j\in \{2,3,...,N\}$ in the multi-vehicle system \ref{sys2}, a distributed controller is proposed as follows to drive to its desired position with maintaining the connectivity and avoiding collisions with other agent $i\in \{1,2,...,N\},i\ne j$:
\begin{equation}
    \begin{array}{rcl}
        u_j&=&\begin{cases}
            \min_{i\in L|J_i<0}u_{j|i},&d_s\le d_{ij}\le R_c,\\
            u_{ij},&R_c\le d_{ij},
        \end{cases}
        \vspace{0.5em}\\
                \omega_j&=& \displaystyle \frac{(B\lambda_j \dot{\phi}_j-\lambda_j u_j\tan{\gamma_j}+B\ddot{\phi}_j)u_j+(\dot{\phi}_j-\lambda_j\theta_j+\lambda_j\phi_j)\dot{u}_j}{u_j^2+B(\dot{\phi}_j-\lambda_j \theta_j+\lambda_j\phi_j)^2},\\
    \end{array}
\end{equation}
where $L\in \{a,b,c,...\}$ is the set of agents that located in the safety region of agent $j$, $J_i=\mathbf{r_{ij}}^T\left[\cos\phi_j\ \sin\phi_j\right]^T$, $\mathbf{r_{ij}}=\mathbf{r_j}-\mathbf{r_i}$, $d_{ij}=||\mathbf{r_{ij}}||$ is the distance between two agents, $u_{j|i}$ and $u_{jc}$ are given as:
\begin{equation}
    \begin{array}{rcl}
         u_{j|i}&=&u_{jc}\displaystyle \frac{d_{ij}-d_s}{R_c-d_s}+u_{js|i} \displaystyle \frac{R_c-d_{ij}}{R_c-d_s},
         \vspace{0.5em}\\
         u_{jc}&=&k_j\tanh{(||\mathbf{r_j}-\mathbf{r_{jd}}||)},
         \vspace{0.5em}\\
         u_{js|i}&=&u_i \displaystyle \frac{\mathbf{r_{ij}}^T\mathbf{\eta_i}}{\mathbf{r_{ij}}^T\mathbf{\eta_j}},
    \end{array}
\end{equation}
in which $\eta_j=\left[\cos\phi_j\ \sin\phi_j\right]^T$, $\phi_j=\text{atan2}\left(\frac{-\partial{V_j}}{-\partial{y_j}},\frac{-\partial{V_j}}{-\partial{x_j}}\right)$ and $k_j>0,\ \eta_j>0$.

\emph{Proof}: We will first prove each agent $i$ converge to its desired position. Let us revisit the developed positive definite Lyapunov-like barrier function $V_i$ derived from Section \uppercase\expandafter{\romannumeral3}. In the rest of this proof, we will check whether $\dot V_i$ is negative such that all the agents can converge to their desired goal positions. Let  $\zeta_i=\left[\frac{\partial V_i}{\partial x_i},\frac{\partial V_i}{\partial y_i}\right]^T$. One can obtain:
\begin{equation}
    \dot V_i=\left[\frac{\partial V_i}{\partial x_i},\frac{\partial V_i}{\partial y_i}\right]\left[\begin{array}{cc}
         \dot x_i\\
         \dot y_i
    \end{array}\right]=\left(\frac{\partial V_i}{\partial x_i}\cos{\theta_i}+\frac{\partial V_i}{\partial y_i}\sin{\theta_i}\right)u_i.
\end{equation}
By exploiting the proposed $u_i$, one has:
\begin{equation}
    \dot V_i=k_i\left(\frac{\partial V_i}{\partial x_i}\cos{\theta_i}+\frac{\partial V_i}{\partial y_i}\sin{\theta_i}\right)\tanh{(||\mathbf{r}_i-\mathbf{r_{1d}}||)}
\end{equation}
Given that $\mathbf{\zeta_i}=\left[\frac{\partial V_i}{\partial x_i}\ \frac{\partial V_i}{\partial y_i} \right]^T$,and $\phi_i=\textrm{atan2}\left(-\frac{\partial V_i}{\partial x_i},-\frac{\partial V_i}{\partial y_i}\right)$, we have:
\begin{equation}
    \frac{\partial V_i}{\partial x_i}=-||\mathbf{\zeta_i}||\cos{\phi_i},\ 
    \frac{\partial V_i}{\partial y_i}=-||\mathbf{\zeta_i}||\sin{\phi_i}.
\end{equation}
It further yields
\begin{equation}
    \dot V_i=-\|\zeta_i\|u_i-\sum^{N}_{k=1,k\neq i} \left( \frac{\zeta_{ik}^T \zeta_k}{\|\zeta_k\|} u_k \right)
    \label{deLyapunov2}
\end{equation}
Given that $\mu_1 \in (0,1)$, $\mu_2 > 1$, $\mu_1 V_i< \| \zeta_i\|$, $\sum^{N}_{k=1, k\neq i} \|\zeta_{ik}\|< \mu_2 V_i$, equation \eqref{deLyapunov2} yields
\begin{equation}
    \dot V_i=-\mu_1 V_i k_i \tanh(\|r_i-r_{id}\|) + \max_{k \neq i} \{u_k\}\mu_2 V_i.
    \label{deLyapunov3}
\end{equation}
Based on the input-to-state stability theory and properties of saddle point \cite{ref9}, it follows that the time derivative of $V_j$ is negative on a compact subset of the concerned constrained set $\mathcal{K}_i$. It thus completes the proofs of convergence to goal destinations.

Next, we will show by using the proposed controller, it ensures the convergence of all agents to their goal destinations. Then, it will be presented that the proposed controller guarantees  the inter-vehicle collision avoidance and connectivity maintenance.  

Regarding the collision avoidance, for agents $i$ and $j$, let us first check the requirement $c_{ij}$ and its derivative:
\begin{equation}
    c_{ij}=(x_i-x_j)^2+(y_i-y_j)^2-d^2_s\geq 0,~k\in\{1,2,\dots,N\}.
\end{equation}
For agents $i$ and $j$, the derivative of $c_ij$ can be computed around the equilibria $\theta_i=\phi_i$ and $\theta_k=\phi_k$ as
\begin{equation}
   \frac{d}{dt}c_{ij}=2u_i \textbf{r}_{ji}^T\left[\begin{array}{cc}
         \cos \phi_i\\
         \sin \phi_i
    \end{array}\right]-2u_j \textbf{r}_{ji}^T\left[\begin{array}{cc}
         \cos \phi_j\\
         \sin \phi_j
    \end{array}\right].
    \label{derivativecij}
\end{equation}
where velocities $u_i$ and $u_j$ are positive, and $\textbf{r}_{ji}=\textbf{r}_i-\textbf{r}_j$. Definite $\textbf{e}_i=[\cos \phi_i, \sin \phi_i]$ and $\textbf{e}_j=[\cos \phi_j, \sin \phi_j]$. It is not difficult to find from \eqref{derivativecij} that collision between agent $i$ and $j$ can be avoided if $\frac{d}{dt}c_{ij} \geq 0$, i.e., the distance between agent $i$ and $j$ is remained or increasing. In order to satisfy this condition, we design the controller in two cases. Case 1: If $2u_j r_{ij}^T e_j \geq 0$, the agent $i$ is approaching to agent $j$ or remains the distance with agent $j$. Provided that the agent $i$ driven with 
\begin{equation}
    u_{jc}=k_j\tanh{(||\mathbf{r_j}-\mathbf{r_{jd}}||)},
\end{equation}
it renders agent $i$ leaving away from agent $j$, which avoids the collision between agent $i$ and agent $j$. Case 2: If $2u_j r_{ij}^T e_j < 0$, the agent $j$ is approaching to agent $i$. Collision between these two agents could be prevented if the second item in the left hand of equation \eqref{derivativecij} is "positive" enough to enable $\frac{d}{dt}c_{ij} >0$. To achieve this, we could let agent j driven by the controller:
\begin{equation}
    u_{j|i}=u_{jc}\displaystyle \frac{d_{ij}-d_s}{R_c-d_s}+u_{js|i} \displaystyle \frac{R_c-d_{ij}}{R_c-d_s},
\end{equation}
in which 
\begin{equation}
    u_{is|j}\leq u_j\frac{\textbf{r}_{ji}^T \textbf{e}_i}{\textbf{r}_{ji}^T \textbf{e}_i}
\end{equation}
In this way, the distance between agent $i$ and agent $j$ can be enlarged and collision could be prevented. Hence, the following distributed controller can be adopted to ensure the collision avoidance for agent $i$:
\begin{equation}
    u^*_{i}= \min_{i\in \mathcal{I}| J_i<0} u_{i|j}.
\end{equation}
where $J_i=\textbf{r}_{ij}\textbf{e}_j$. 

Similarly, for connectivity maintenance, we consider the requirement: $c_{i0}=R^2-\|r_j-r_0\|$ focusing on the boundary of the set $\mathcal{K}_i$. Here we also compute the derivative of this constraint:
\begin{equation}
   \frac{d}{dt}c_{i0}=-2u_i \frac{r_{0j}^T e_i}{\|r_{0i}\|},
    \label{derivativeci0}
\end{equation}
where $r_{0i}=r_i-r_0$. One can observe that either agent $i$ is in conflict with no agent or with at least one agent $j$, the condition $r_{0j}^T e_i \leq 0$ always holds under the connectivity maintenance constraint. It thus completes the proof. \hfill $\square$

\subsection{Multi-vehicle Systems with Misbehaving Agents}
In a bicycle system with malfunctioning agents, we assume these agents whose distributed controller is not functioning or working properly. As a result, these agents move randomly within the working space. However, they are still able to transmit information to other agents. In other words, controllable agents can sense the position, speed, and orientation of uncontrollable agents. On the other hand, uncontrollable agents are unable to receive any information from other agents. Consequently, they lack a controller and are unable to avoid collisions or maintain connectivity. It is reasonable to assume that the movement trajectories of all uncontrollable agents do not intersect with the destinations of controllable agents

For leader agent $j=1$, in normal system, leader agent do not avoid collision with any other follower agents, converging to its desired position, which may collide with uncontrollable agents. In this case, those uncontrollable agents should be considered as physical obstacles in working space. The minimum distance between agents and those b=obstacles can be defined as the minimum distance between two different agents $d_s$.

In this way, for follower agents $j\in \{2,3,...,N\}$, they should consider both other agents $i\in\{1,2,...,N\},i\ne j$ and physical obstacles (uncontrollable agents).

\begin{figure}[ht]
\begin{minipage}[t]{0.3\textwidth}
\includegraphics[width=\textwidth]{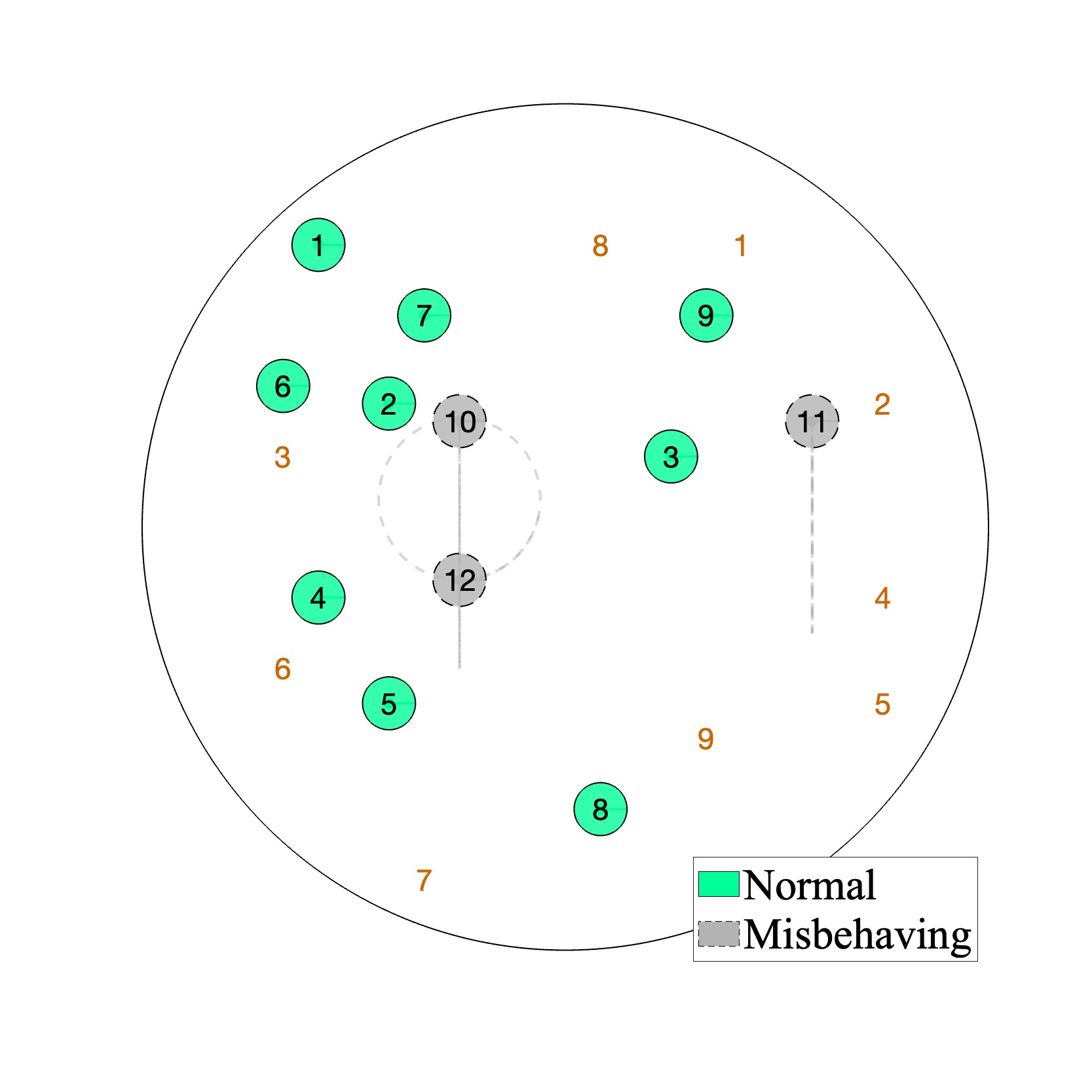}
\begin{center}
    \textit{t=0.0 sec}\\
\end{center}
\end{minipage}
\begin{minipage}[t]{0.3\textwidth}
\includegraphics[width=\textwidth]{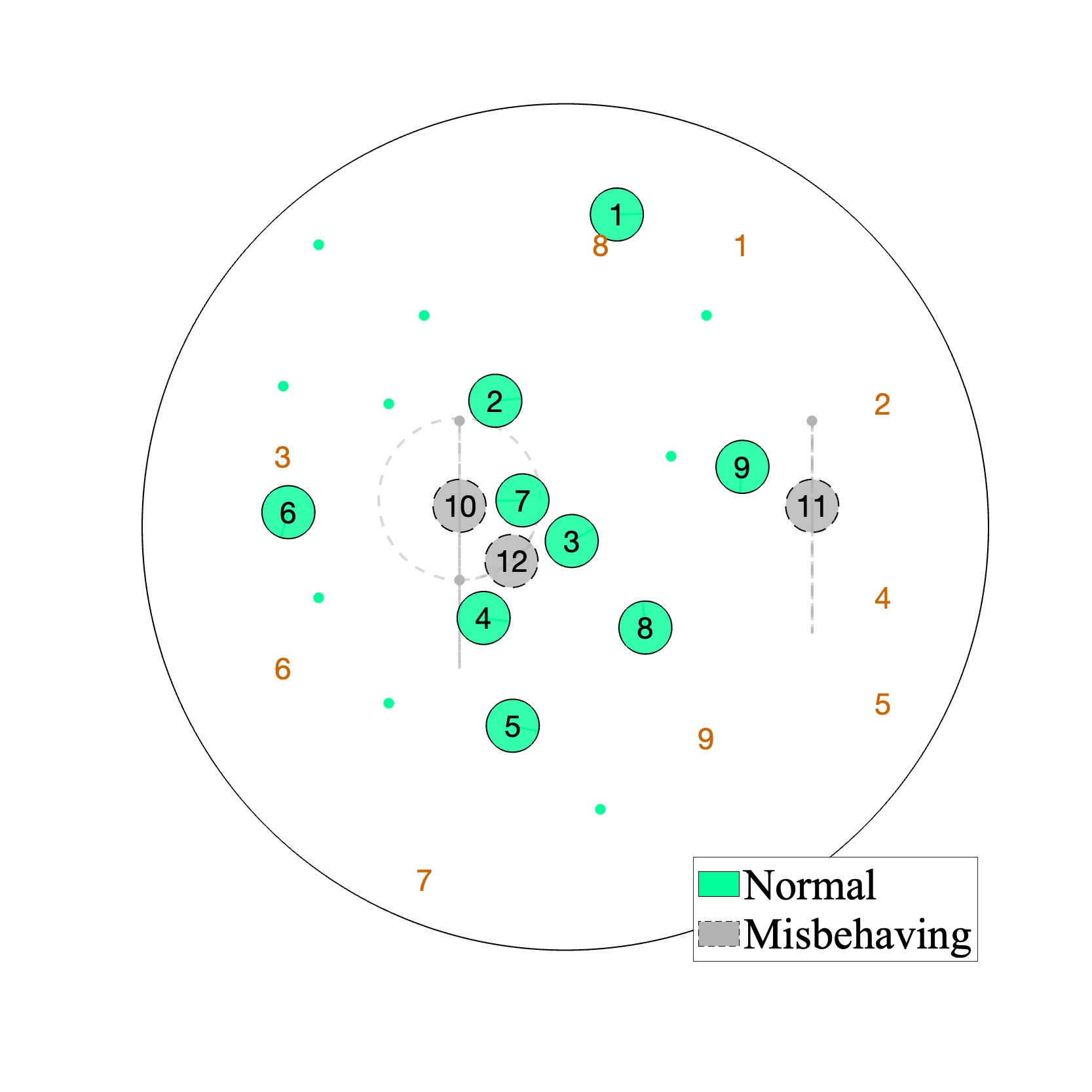}
\begin{center}
    \textit{t=2.0 sec}\\
\end{center}
\end{minipage}
\begin{minipage}[t]{0.3\textwidth}
\includegraphics[width=\textwidth]{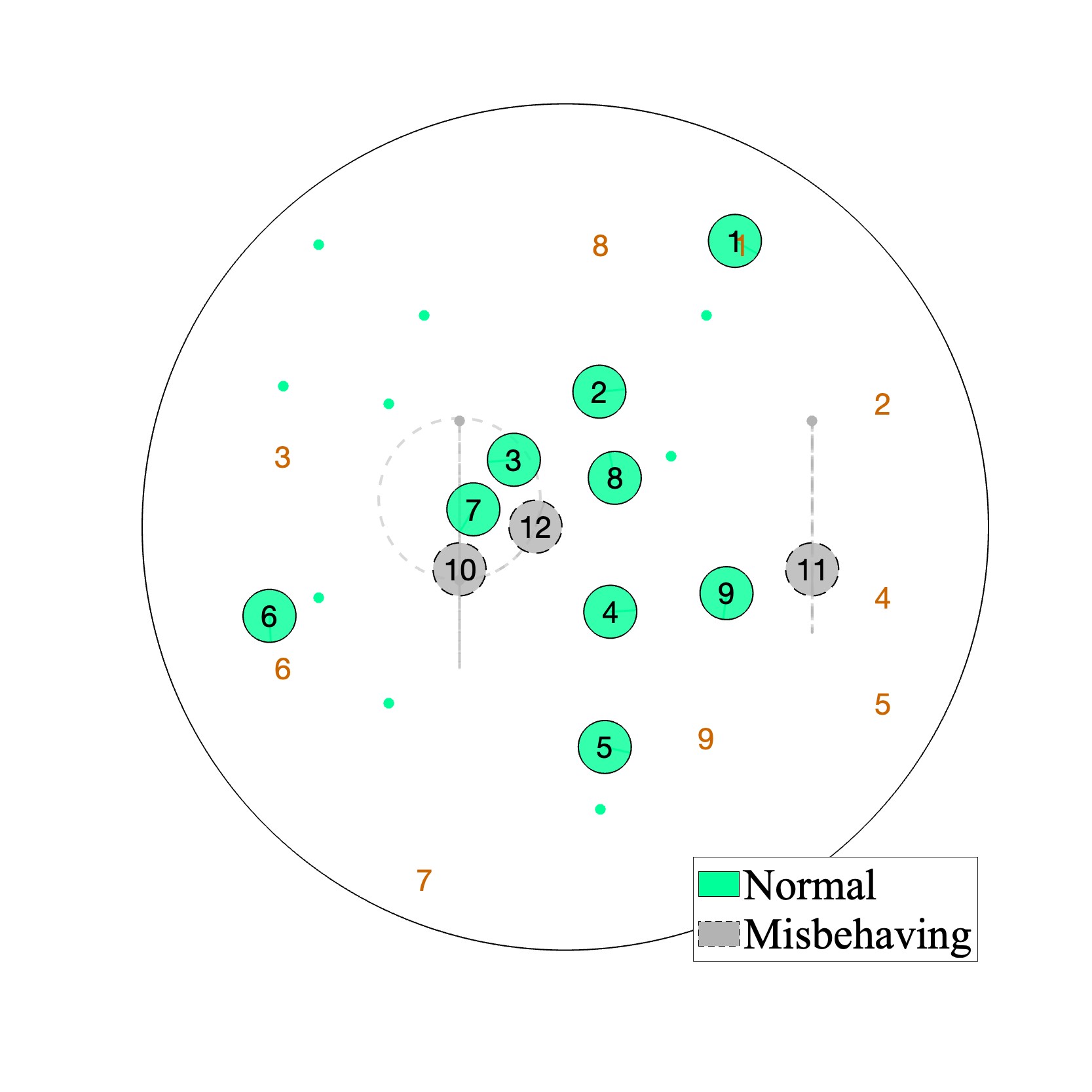}
\begin{center}
    \textit{t=3.5 sec}\\
\end{center}
\end{minipage}
\vspace{0.5em}\\

\begin{minipage}[t]{0.3\textwidth}
\includegraphics[width=\textwidth]{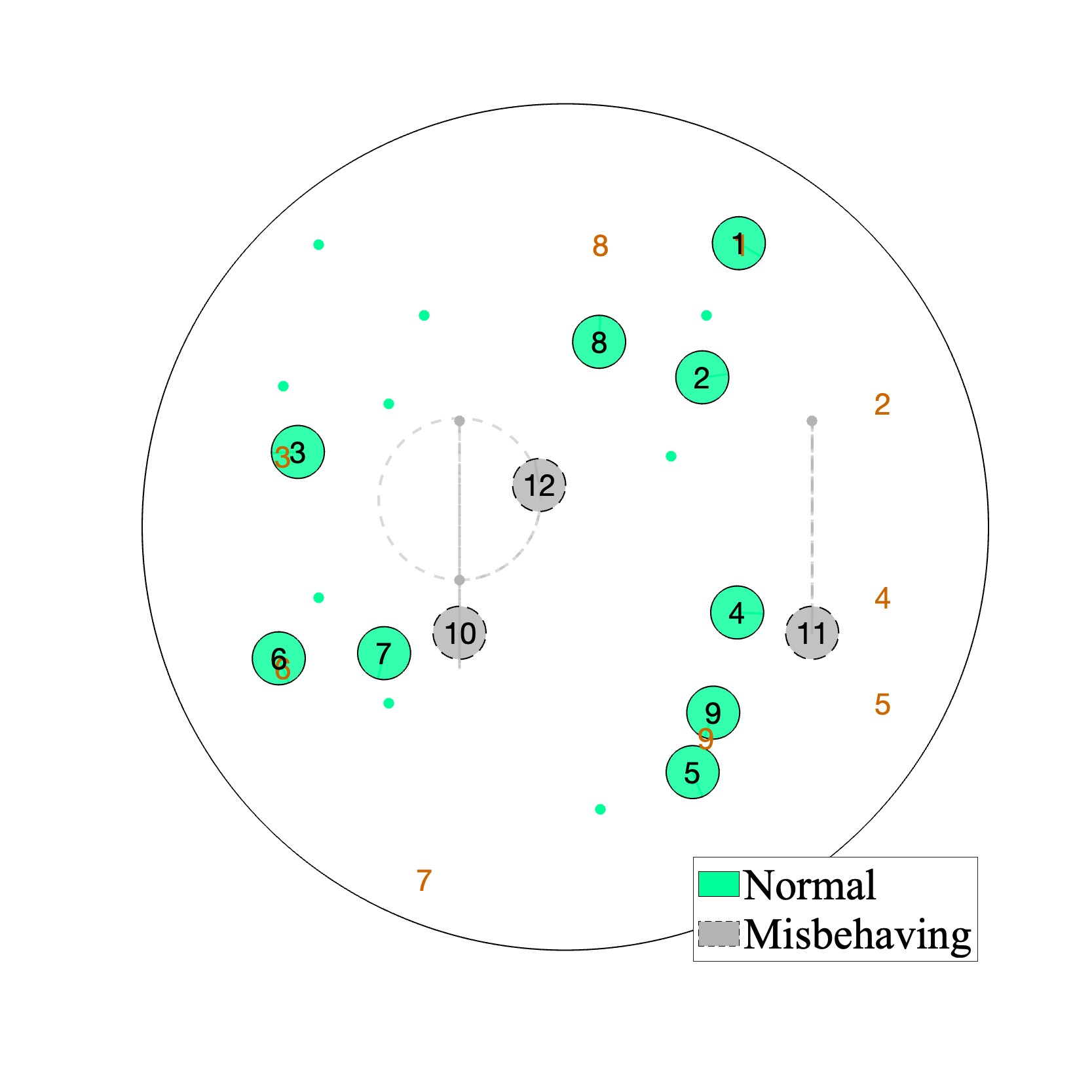}
\begin{center}
    \textit{t=5.0 sec}\\
\end{center}
\end{minipage}
\begin{minipage}[t]{0.3\textwidth}
\includegraphics[width=\textwidth]{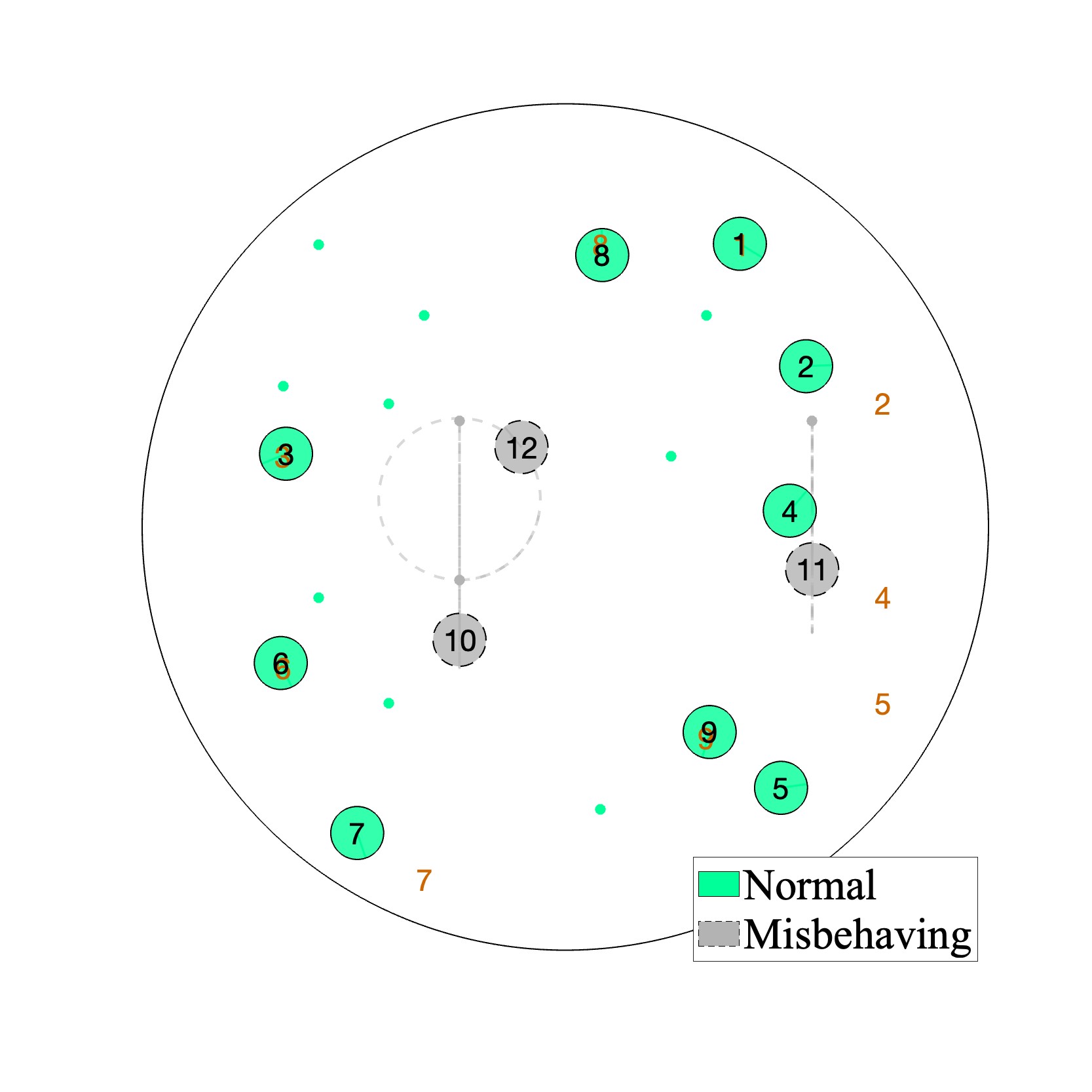}
\begin{center}
    \textit{t=6.5 sec}\\
\end{center}
\end{minipage}
\begin{minipage}[t]{0.3\textwidth}
\includegraphics[width=\textwidth]{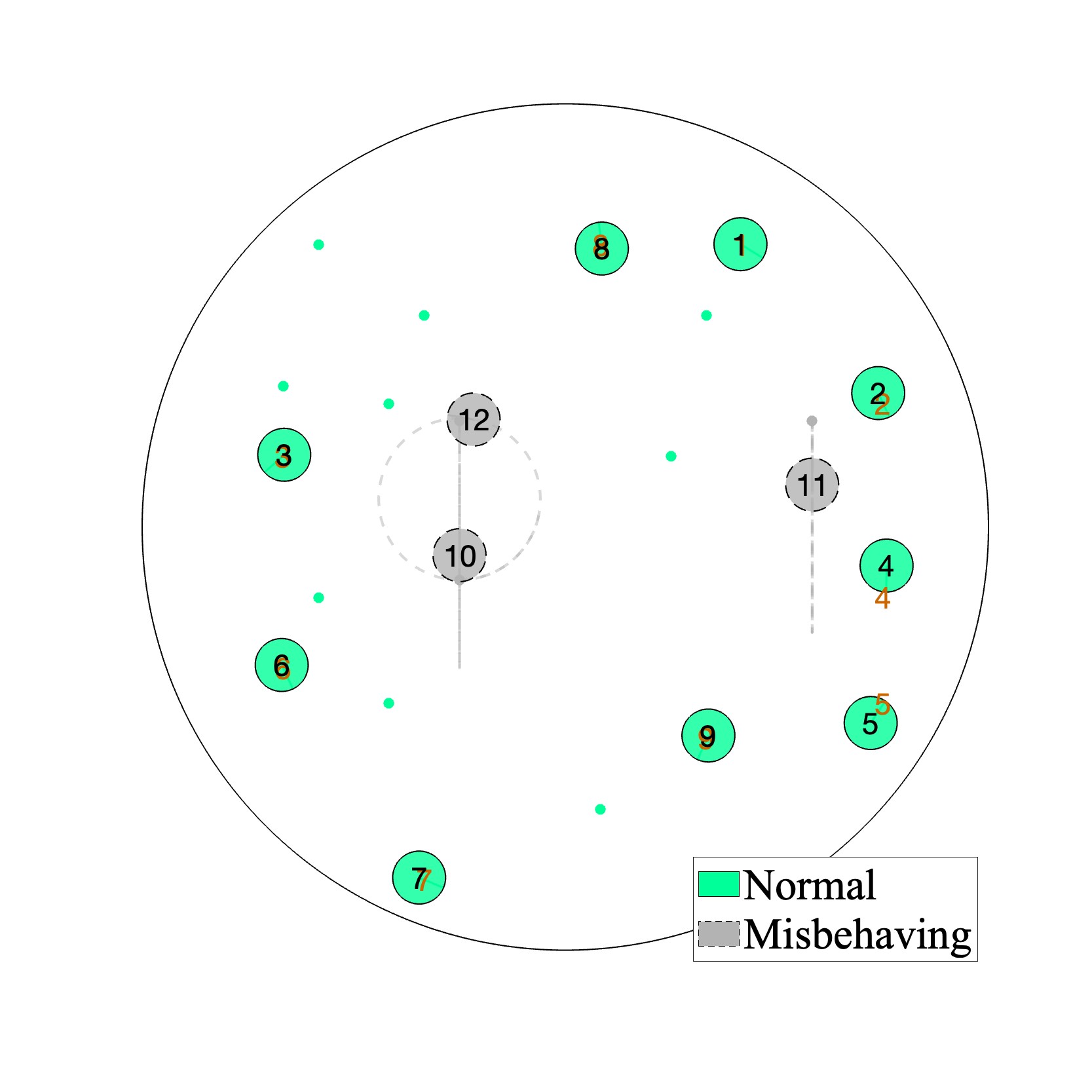}
\begin{center}
    \textit{t=8.5 sec}\\
\end{center}
\end{minipage}
\vspace{0.5em}\\

\begin{minipage}[t]{0.3\textwidth}
\includegraphics[width=\textwidth]{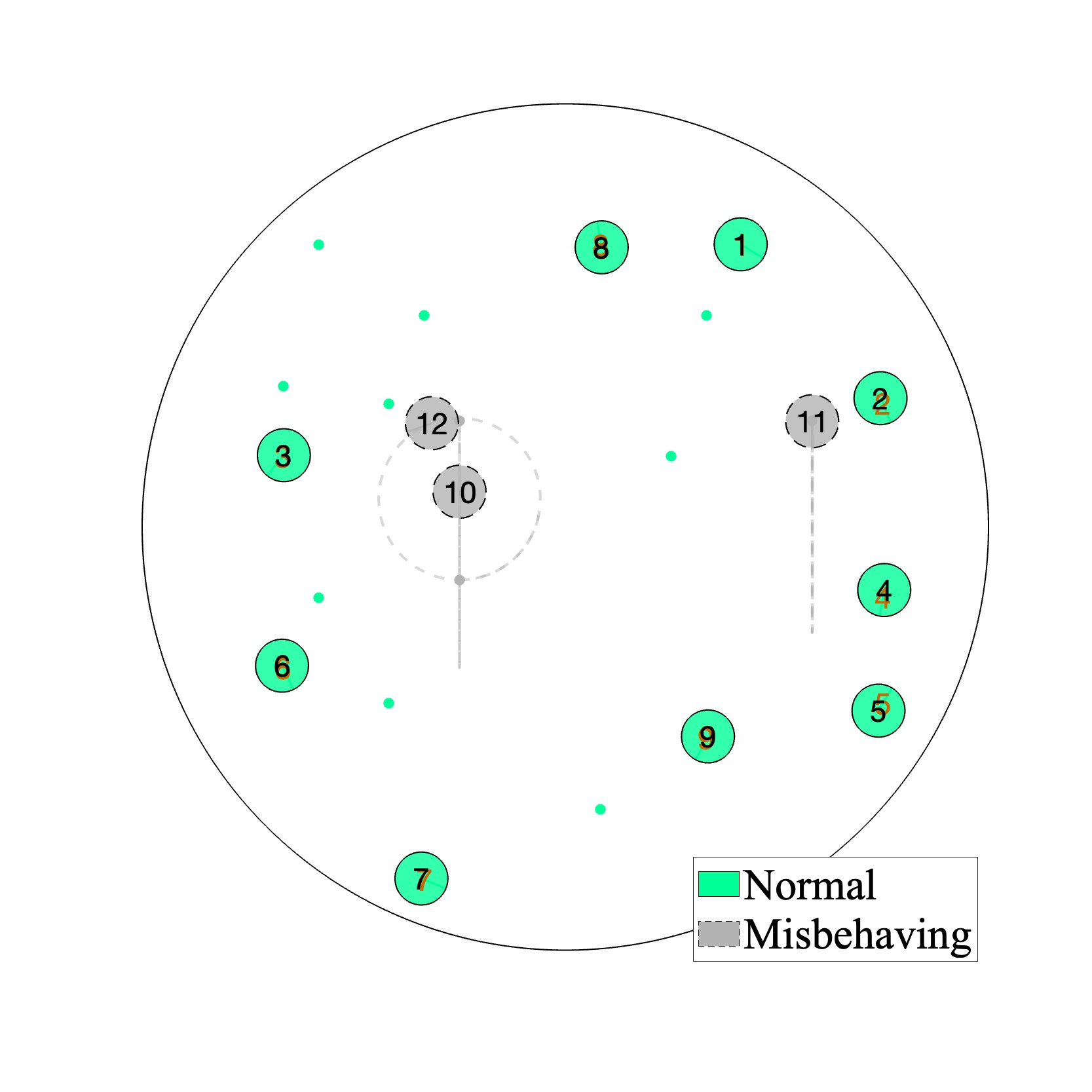}
\begin{center}
    \textit{t=10.0 sec}\\
\end{center}
\end{minipage}
\begin{minipage}[t]{0.3\textwidth}
\includegraphics[width=\textwidth]{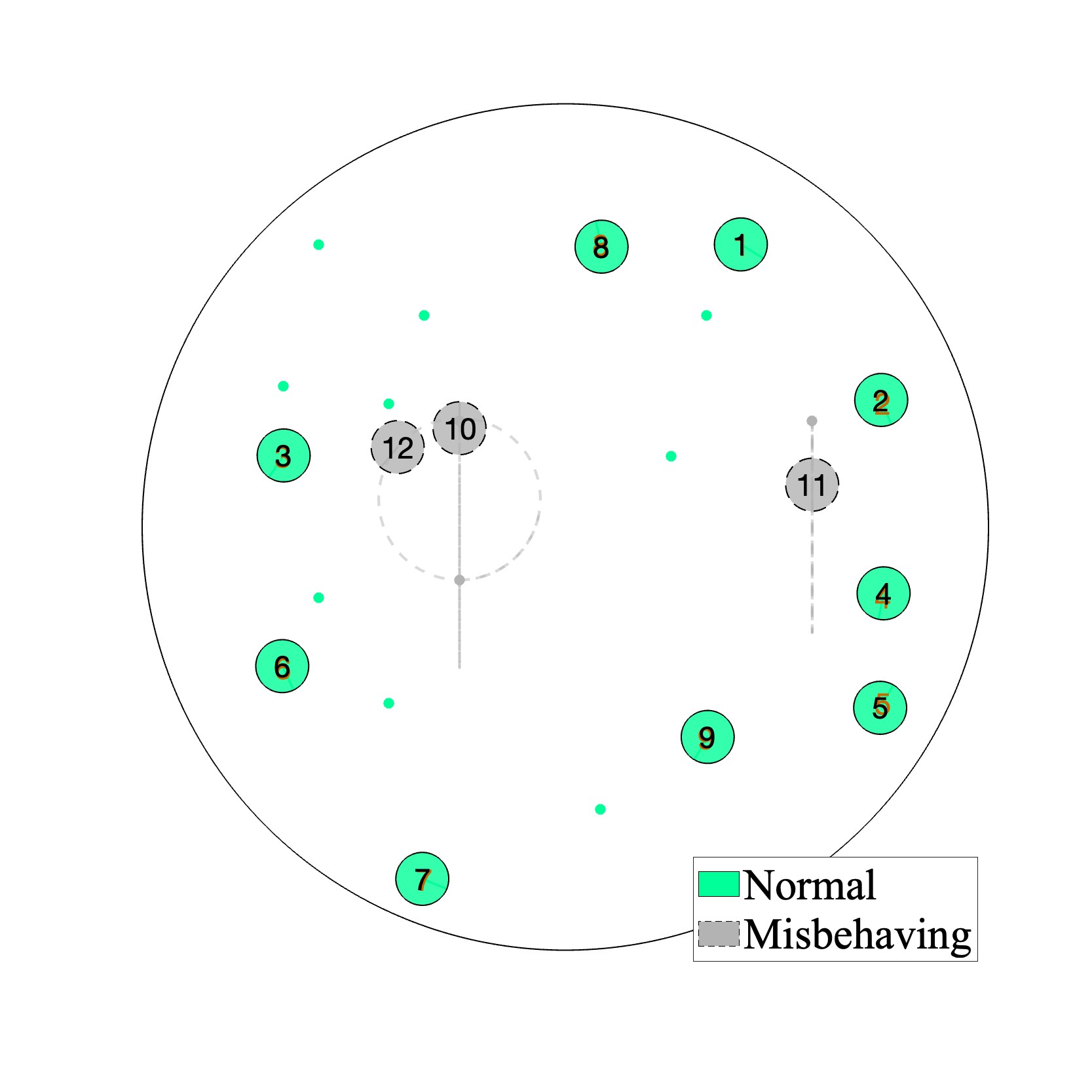}
\begin{center}
    \textit{t=11.5 sec}\\
\end{center}
\end{minipage}
\begin{minipage}[t]{0.3\textwidth}
\includegraphics[width=\textwidth]{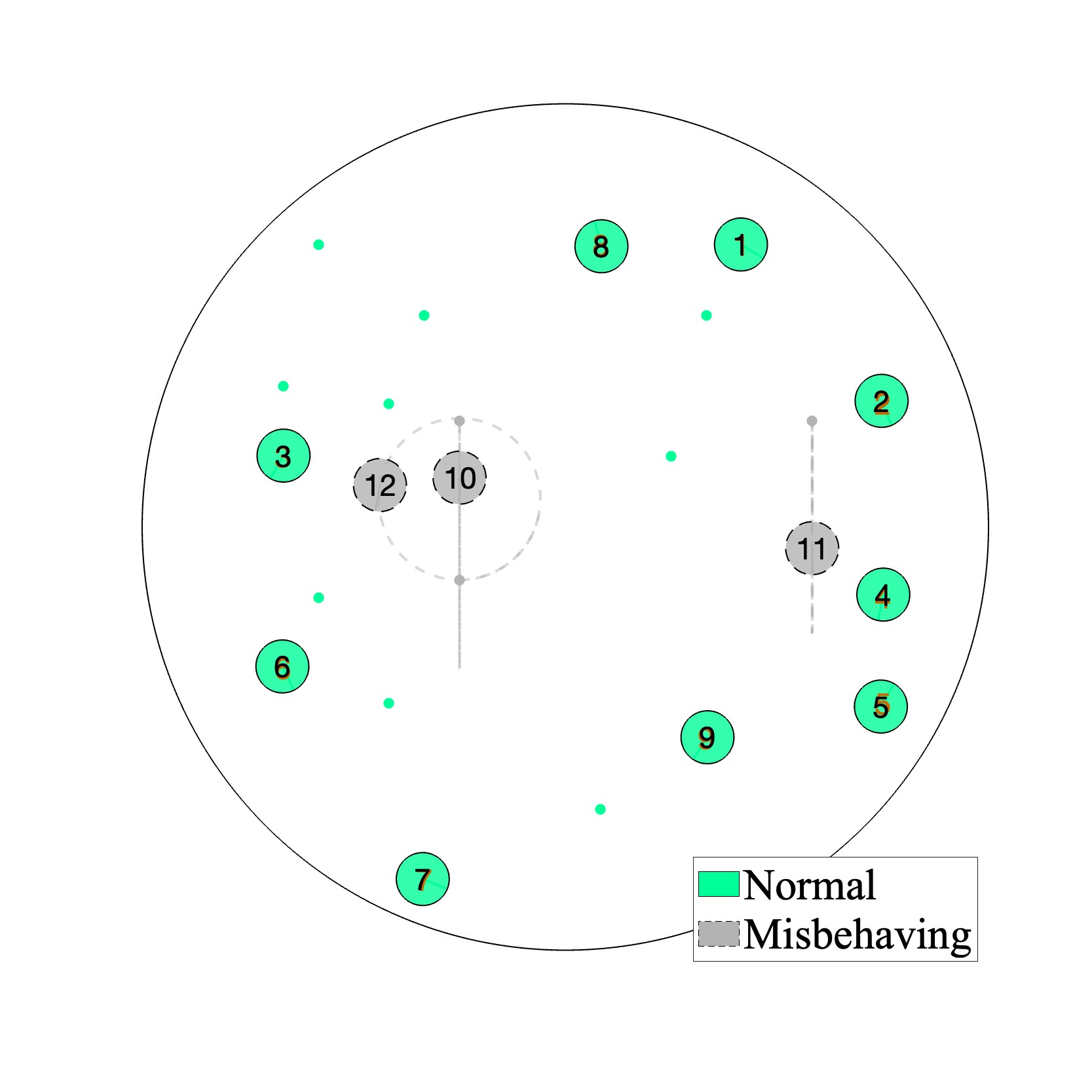}
\begin{center}
    \textit{t=13.0 sec}\\
\end{center}
\end{minipage}
\caption{The motion of multi-vehicle in the presence of misbehaving vehicles (white agents).}
\end{figure}

The controller of the leader agent $j=1$ in this case adopts the collision avoidance part in the previous subsection, but only considers other uncontrollable agents $i$ within its sensing region. The controller can be defined as:
$$
    \begin{array}{rcl}
        u_j&=&\begin{cases}
            \min_{i\in L|J_i<0}u_{j|i},&d_s\le d_{ij}\le R_c\\
            u_{ij},&R_c\le d_{ij}
        \end{cases}
        \vspace{0.5em}\\
                        \omega_j&=& \displaystyle \frac{(B\lambda_j \dot{\phi}_j-\lambda_j u_j\tan{\gamma_j}+B\ddot{\phi}_j)u_j+(\dot{\phi}_j-\lambda_j\theta_j+\lambda_j\phi_j)\dot{u}_j}{u_j^2+B(\dot{\phi}_j-\lambda_j \theta_j+\lambda_j\phi_j)^2},\\
    \end{array}
$$
where for the leader agent $j=1$, $i\in\{\text{agents with sensing region}\}\cap \{\text{uncontrollable agents}\}$, and for follower agents $j\in \{2,3,...,N\}$, $i\in\{\text{agents with sensing region}\}$.

\section{Simulations}

The efficacy of this controller can be demonstrated through MATLAB. In simulation, we set the radius of agent $r_a=0.75m$, the length of bicycle $B=0.25m$ and the connectivity region O $R_0=12m$ centered at $\mathbf{r_0}=[0\ 0]^T$, where the center can be modified by users. Then, the sensing region, safety region and collision avoidance region are chosen respectively as $R_S=1.80m, Rz=1.60m, R_c=2.25\times r_a$. And we tried some different $\lambda_i$, when $\lambda_i=2.3$, simulations have relatively stable performance. We have tried many different cases, and here we choose to show some of the most representative ones. The purpose of this case is to test whether the system can remain stable when some agents are out of control.

\begin{figure}[h]
\centering
\includegraphics[width=0.7\linewidth]{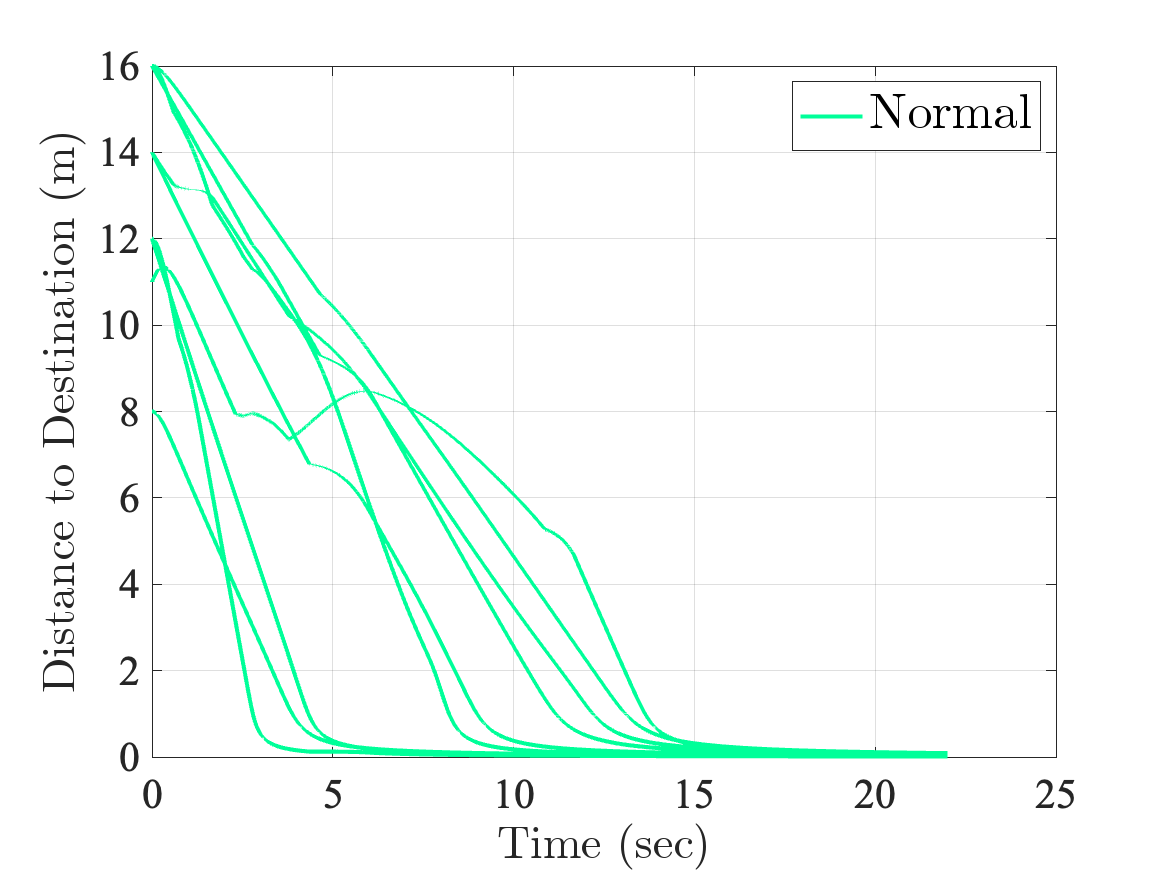}
\caption{Distance of Each Normal Agent to Their Destination Over Time.}
\label{fig:convert_to_destination}
\label{rij}
\end{figure}

The nine figures above show the trajectory of convergence of 12 agents with 3 of them being uncontrollable (agents 10, 11, and 12). Agents 10 and 11 are repeatably moving between two fixed points, while agent 12 has a circular trajectory of movement. We can observe that under the proposed controller, all 9 controllable agents can reach their desired position (solid dots of the corresponding colors) in 13 seconds, as shown in Figure \ref{fig:convert_to_destination}.

\begin{figure}[h]
\centering
\includegraphics[width=0.7\linewidth]{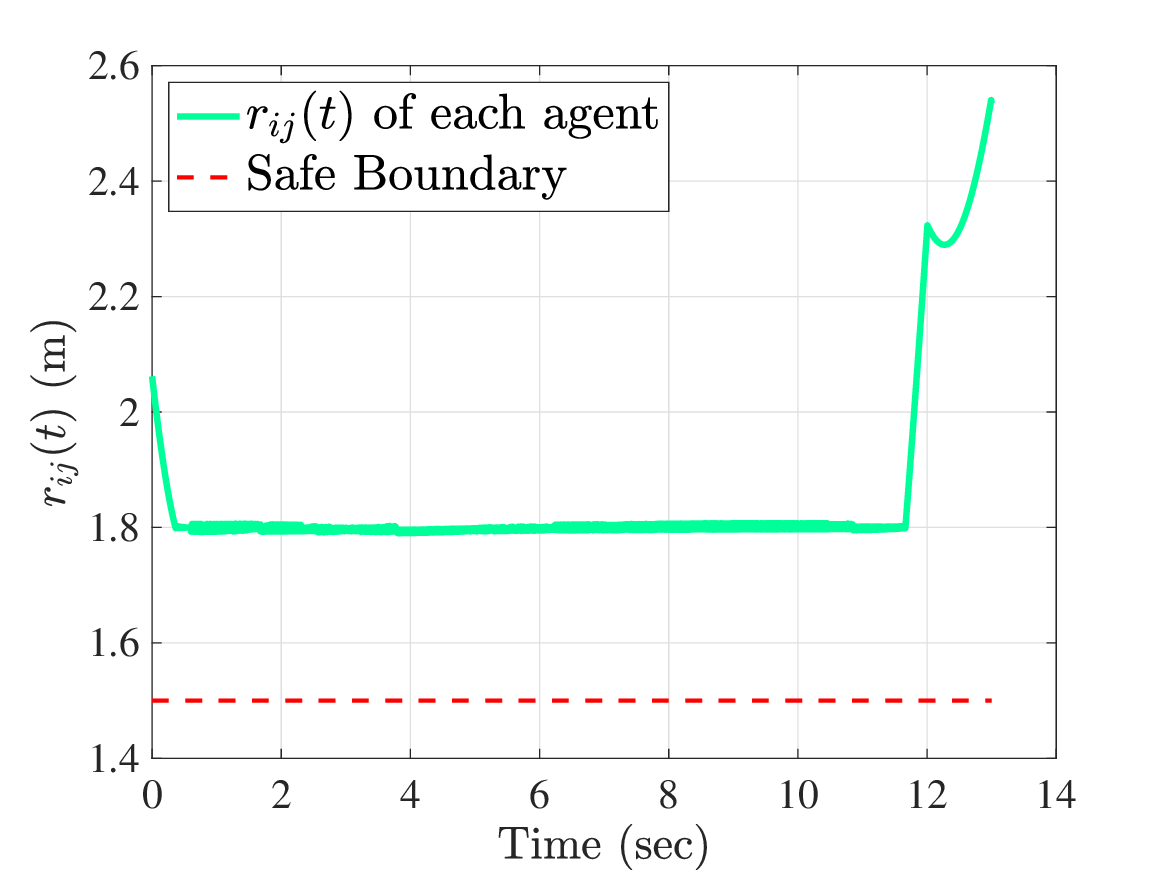}
\caption{The evolution of the inter-agent distances $\|r_{ij}(t)\|$.}
\label{rij}
\end{figure}

% \begin{figure}[h]
% \centering
% \includegraphics[width=0.7\linewidth]{figures/r_ij.jpg}
% \caption{The evolution of the inter-agent distances $\|r_{ij}(t)\|$.}
% \label{rij}
% \end{figure}

Here we show the evolution of the distances between agents $\|r_{ij}\|(t)$ in Figure \ref{rij}. We can see that due to the uncontrollable agents, some of the inter-agent distances are kept changed regularly. And we have simulated some cases with all agents controllable, in which the evolution of the inter-agent distances $d_{ij}(t)$ would finally remain stable as depicted in Figure \ref{fig:convert_to_destination}. In some cases, there are too many agents crowded into a small space. Some agents may sometimes jam together for a long time.

\section{Conclusion}

In this work, a distributed controller for bicycle modeled leader-follower multi-vehicle systems is designed based on constructing a unified Lyapunov-like barrier function. Compared to the results based on unicycle modeled systems, this work proposes a distributed controller for a more practical model thus more applicable to real engineering practices. Future efforts could devote to finite time stability, which may have additional requirements on Lyapunov functions \cite{ref6}.  We are also very interested to consider the uncertainties and disturbances in sensing and measurement. Constructing the Lyapunov-like barrier function usually is not easy in controller design, thus automatic ways to generate barrier functions will also be investigated, like methods by Quadratic Programming \cite{ref8} and Sum-of-Squares method \cite{wang18ACC,han19tac,han12tii}.

\section{Acknowledgement}       
 
Financial support from the Teaching Development and Language Enhancement Grant (4170989), Hong Kong SAR, is gratefully acknowledged. The authors are also grateful for the discussions and valuable inputs from Prof. Jie Huang and Prof. Benmei Chen at Department of Mechanical and Automation Engineering, The Chinese University of Hong Kong.

% References
\section*{References}
\bibliography{acmae} % bibliography data in report.bib
\bibliographystyle{spiebib} % makes bibtex use spiebib.bst

\end{document}